\documentclass[final,authoryear,5p,times,twocolumn]{elsarticle}

\usepackage{amsmath}
\usepackage{amssymb}
\usepackage{fixltx2e}
\usepackage{multirow}

\usepackage{lineno,hyperref}
\modulolinenumbers[5]

\journal{Journal of \LaTeX\ Templates}

\biboptions{authoryear}

\usepackage{color}

\newcommand{\progname}[1]{{\fontfamily{pcr}\selectfont #1}}

\begin{document}

\begin{frontmatter}

\title{Estimation of the height of the first interaction in gamma-ray showers observed by Cherenkov telescopes}

\author[afil]{Julian Sitarek}
\ead{jsitarek@uni.lodz.pl}
\author[afil]{Dorota Sobczy\'nska}
\ead{dsobczynska@uni.lodz.pl}
\author[afil]{Katarzyna Adamczyk}
\ead{kadamczyk@uni.lodz.pl}
\author[afil2]{Micha\l\ Szanecki}
\ead{mitsza@camk.edu.pl}
\author[afil3]{Konrad Bernl\"ohr}
\ead{Konrad.Bernloehr@mpi-hd.mpg.de}

\address[afil]{Department of Astrophysics, The University of \L\'od\'z, ul. Pomorska 149/153, 90-236 \L\'od\'z, Poland}
\address[afil2]{CAMK, ul. Bartycka 18, 00-716 Warsaw, Poland}
\address[afil3]{Max-Planck-Institut f\"ur Kernphysik, P.O. Box 103980, D-69029 Heidelberg, Germany}

\begin{abstract}
  Very high energy gamma rays entering the atmosphere initiate Extensive Air Showers (EAS).
  The Cherenkov light induced by an EAS can be observed by ground-based telescopes to study the primary gamma rays.
  An important parameter of an EAS, determining its evolution, is the height of the first interaction of the primary particle.
  However, this variable cannot be directly measured by Cherenkov telescopes. 
  We study two simple, independent methods for the estimation of the first interaction height.
  We test the methods using the Monte Carlo simulations for the 4 Large Size Telescopes (LST) that are part of the currently constructed Cherenkov Telescope Array (CTA) Observatory. 
  We find that using such an estimated parameter in the gamma/hadron separation can bring a mild improvement ($\sim10-20\%$) in the sensitivity in the energy range $\sim30-200$\,GeV.
\end{abstract}


\end{frontmatter}


\section{Introduction}
Gamma rays entering the Earth's atmosphere interact with the atmospheric nuclei creating cascades of secondary particles, dubbed Extensive Air Showers (EAS). 
Ultrarelativistic particles produced in EAS stimulate generation of short and faint flashes of Cherenkov light, which can be observed by ground-based Imaging Atmospheric Cherenkov Telescopes (IACT). 
The Cherenkov-light images of the showers are used to reconstruct the arrival direction and energy of the primary particle. 
However, EAS produced by primary cosmic rays are over three orders of magnitude more abundant than the ones produced by gamma rays.
Thus, the technique requires an effective method of the elimination of background events from the recorded data. 
The rejection is based on both the directional information \citep{kr98}, stereo reconstruction of the geometry of the shower \citep{ah97} and on the shape of individual images \citep{hi85}. 
An important parameter determining the properties of an EAS is the height at which the first interaction occurred  \citep[see e.g.][]{ma05}.
For a given zenith angle and atmospheric profile this altitude corresponds to an atmospheric depth that primary particle traversed before the first interaction.
It is not used in the classical Hillas-parameters-based analysis of IACTs, however it is one of the fit parameters in model analysis \citep[see e.g.][]{nr09}.
The showers starting earlier or later in the atmosphere have the maximum respectively shallower or deeper in the atmosphere.
The first interaction height affects both the amount of light registered by the telescopes \citep[see e.g.][]{so09a} and the shape of the obtained images. 
Moreover, due to different cross sections for hadronic pion production and electromagnetic pair production, gamma rays have on average its first interaction higher than protons. 
Unfortunately, there is no direct way for a Cherenkov telescope to measure the height of the first interaction of an EAS. 
A useful proxy of it is however given by the reconstructed height of the shower maximum, which helps to remove muon events \citep[see e.g.][]{ah97, al12} and single electromagnetic subcascase background events \citep{si18}. 

The Cherenkov Telescope Array (CTA) is an upcoming observatory consisting of two large arrays of Cherenkov telescopes located in the Southern and Northern hemispheres \citep{acha13}. 
The arrays will be composed of three different types of telescopes: LST, MST, and SST i.e. large (diameter of 23\,m), medium (12\,m), and small (4\,m) size telescopes respectively) to be able to cover the energy range from $\sim 20$\,GeV up to beyond $\sim 300$\,TeV.
The expected excellent performance
of CTA will make the future observations performed with this instrument, the main driver of the gamma-ray astrophysics in the next years \citep{acha17}. 

We study two independent methods for the estimation of the first interaction height that can be applied for arrays of Cherenkov telescopes. 
The methods are tested on a simulated array of four LST telescopes, which are crucial for the performance of CTA at the lowest energies. 
In Section~\ref{sec:sim} we describe the Monte Carlo simulations. 
In Section~\ref{sec:methods} we present the two methods for estimation of the height of the first interaction. 
The application of such parameters in the analysis for separation of gamma rays from the background is studied in Section~\ref{sec:res}. 
The conclusions are given in Section~\ref{sec:con}

\section{Simulation setup}\label{sec:sim}

The simulation data used for this investigation were extracted from the CTA \emph{prod-3} data set for the layout optimization of the CTA-North (La Palma) site.
The original simulation include nine potential LST and 91 potential MST locations. 
Air showers were simulated with \progname{CORSIKA} \citep{he98}, version 6.990 with the \progname{QGSJET-II-03} \citep{ost06a,ost06b,os07} interaction model for high-energy hadronic interactions and \progname{URQMD} \citep{ba98} for low energies ($<80$\,GeV).
The telescope simulation was using \progname{sim\_telarray} \citep{be08}. 

The primary particles were simulated to originate from (around) 20$^\circ$ zenith angle, North and South\footnote{Throughout the paper we use the \progname{CORSIKA} definition of Azimuth, i.e. $0^\circ$ means that the shower is moving towards North  and the telescope is pointing South} due to the different impact of the geomagnetic field (GF).
We use simulations of gamma rays originating from the direction corresponding to the center of the camera.
Their core offsets calculated from the center of the array were simulated to be spread up to 1600\,m.
Diffuse electrons and protons at up to 10$^\circ$ from the center of the camera were simulated with core offsets up to 2100\,m.
The simulated energy range of gamma rays and electrons is 3\,GeV to 330\,TeV, and for protons: 4\,GeV to 660\,TeV, with a spectral slope of $-2$.
In total 84M, 1260M and 130M distinct gamma ray, proton and electron showers respectively were simulated.
To further improve the statistics the showers have been reused 10 times (for gamma rays) or 20 times (for protons and electrons), by randomly shifting their impact point.
Due to the large simulated maximum core offset (and for diffuse particles also the maximum angular distance to the camera center) the real reusage factor of reconstructed events is much smaller. 
  For gamma rays each Corsika-generated shower is used on average between 1.08 (at 30\,GeV) and 1.6 (at 10\,TeV) times, while for protons the corresponding numbers are 1.05 (at 100\,GeV) and 1.4 (at 30\,TeV).
  Therefore we conclude that the reusage of showers does not have an important  effect on the estimation of the statistical uncertainties. 
It should be noted that MC simulations of Cherenkov radiation from electrons with the \progname{CORSIKA} program (including the version used in the \emph{prod-3} production) have an intrinsic simplification.
The Cherenkov radiation is not simulated until the first interaction of the electron occurs.
However, as electrons produce constantly soft Bremsstrahlung, the first interaction registered by the \progname{CORSIKA} program is always very high, typically $\sim$47\,km a.s.l., well above the height where the actual shower starts. 
Hence the lacking Cherenkov photons above this altitude in the simulated data may have only a small effect on our study.
For this study the data corresponding only to stereo-triggered events of the four LSTs was extracted.
The array of 4 LSTs corresponds to the initial approved CTA-North layout, a quadrangle close to a square with a 100\,m side at a mean altitude of 2180\,m.
The LST telescope model in these simulations is identical to the one in the later \emph{prod-3b} \citep{ma17} simulations, although LST positions were slightly modified. 

The extraction of signal amplitudes from simulated waveforms, image cleaning and its parametrization, gamma/hadron separation and energy estimation is done using \progname{MARS/Chimp} chain \citep{za13, al16b, si18}.
In particular, we used two-pass core-boundary pixel cleaning.
The first pass is done with thresholds 6 and 3 photoelectrons (phe) for core and boundary pixels, respectively.
The pixels from the first pass are used to calculate the  time gradient measured along the main axis of the image.
This allows us to reextract the signals from a narrower search window and do a second cleaning pass with lower thresholds 4 -- 2 phe.
The standard \progname{MARS/Chimp} chain stereo-reconstruction method is described in detail in \citep{si18}. 
For the stereo reconstruction only information from telescopes with total signal of above 50 phe in the camera and distance of the center of the gravity within the 80\% of the camera radius are used.
In order to include the effect of the GF we analyze the MC simulations generated in two opposite direction of the azimuth angle  (i.e. $0^\circ$ and $180^\circ$) separately.

\section{Estimation of the height of the first interaction}\label{sec:methods}
We derived two methods to estimate the height in the atmosphere at which the conversion of the primary gamma ray into an $e^+e^-$ pair happened. 
Both methods are independent, have different limitations and in principle can be combined to obtain a more reliable estimation. 

\subsection{Using estimated energy and shower maximum}\label{sec:hmax}
\begin{figure}[t]
\includegraphics[width=0.49\textwidth]{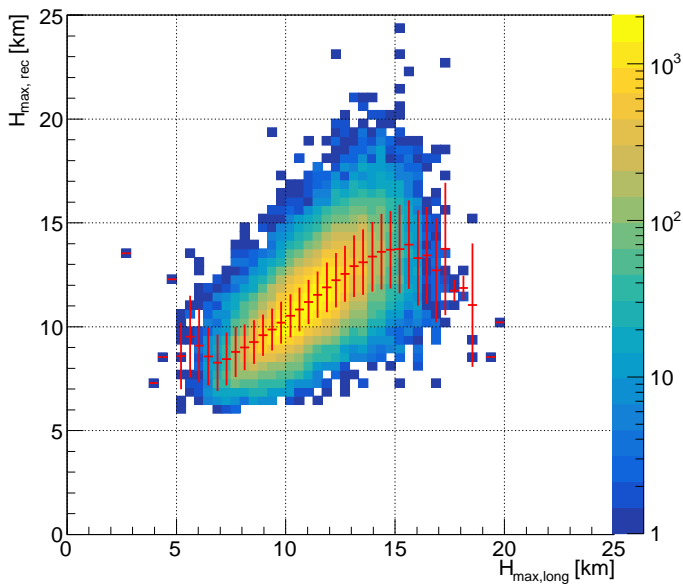}
\caption{
  Height of the shower maximum reconstructed from the Cherenkov images of the event as a function of the height of the shower maximum obtained from the \progname{CORSIKA} fit to the longitudinal distribution of charged particles. 
  Gamma-like events with $\theta<0.1^\circ$, that were obtained from the simulated sample of the primary gamma ray at $180^\circ$ azimuth, are plotted in this Figure.
  The red lines show the average and RMS of the reconstructed height of the shower maximum as a function of the \progname{CORSIKA} fit value.
}
\label{fig:hmaxrec}
\end{figure}

Contrary to the height of the first interaction, the altitude of the shower maximum can be estimated in a rather straight-forward way using geometrical 3D reconstruction of the shower \citep{ah97}.
In fact the height of the shower maximum carries a strong imprint of the first interaction. 
As an example, the fluctuations of the shower maximum for a 100\,GeV $\gamma$ ray are about $60\,\mathrm{g\,cm^{-2}}$ \citep[see e.g.][]{si18}, most of them are caused by the fluctuations of the first interaction point $9/7\,X_{0}\approx47\,\mathrm{g\,cm^{-2}}$, where $X_0$ is the radiation length in air.
The depth of the shower maximum (and the longitudal distribution of the shower in general) is also used by the cosmic ray experiments for the measurement of the proton-air cross-section, that is directly related to the height of the first interaction \citep[see e.g.][]{el82,pao12}. 
The average depth of the shower maximum depends logarithmically on the energy of the shower.
Note, that both the atmospheric absorption and the limited field of view of the telescopes can shift the observed maximum of Cherenkov radiation (by $\sim20\,\mathrm{g\,cm^{-2}}$ in the energy range 20 -- 500\,GeV, \citealp{so09a}). 
The value of the reconstructed height of the shower maximum is compared to the value obtained from the \progname{CORSIKA} longitudinal development of charged particles fit in Fig.~\ref{fig:hmaxrec}.
For the bulk of the showers with the maximum at the height of $\sim 11$\,km there is nearly no bias in the estimation.
  Showers with their maximum  very high in the atmosphere ($\gtrsim 16$\,km) are reconstructed at lower altitudes, most probably due to absorption of the Cherenkov light from the top of the shower and higher energy threshold for production of Cherenkov light at this height. 
  On the other hand showers with their maximum deep in the atmosphere ($\lesssim 7$\,km) are reconstructed preferentially at a larger height.
  The bottom part of such showers will often not be visible by the telescope due to too large angle between the shower axis and the direction to the telescope, causing such bias in the reconstruction of the shower maximum.

Using the estimated energy of the shower, $E_{est}$, we calculate the number of shower generations from the first interaction up to maximum as $N_{\mathrm{int}}=\log(E_{est}/\mathrm{80\,MeV}) - 9/7$. 
If the shower maximum is at thickness $X_\mathrm{max}$, the corresponding thickness of the first interaction will be $X_{\mathrm{1st}}=X_\mathrm{max}-N_{\mathrm{int}}\times X_{0}$. 
Using the atmospheric profile assumed in the simulations, this thickness is converted into the height of the first interaction.
Note that due to fluctuations it is possible that $X_{\mathrm{1st}}$ is very small, or even negative. 
In such a case we store a generic high number as the height of the first interaction (in those calculations 60\,km is used). 
For primary gamma rays with energies $\lesssim 100$\,GeV we have obtained $\lesssim35\%$ of such events. 

\begin{figure}[t]
\includegraphics[width=0.49\textwidth]{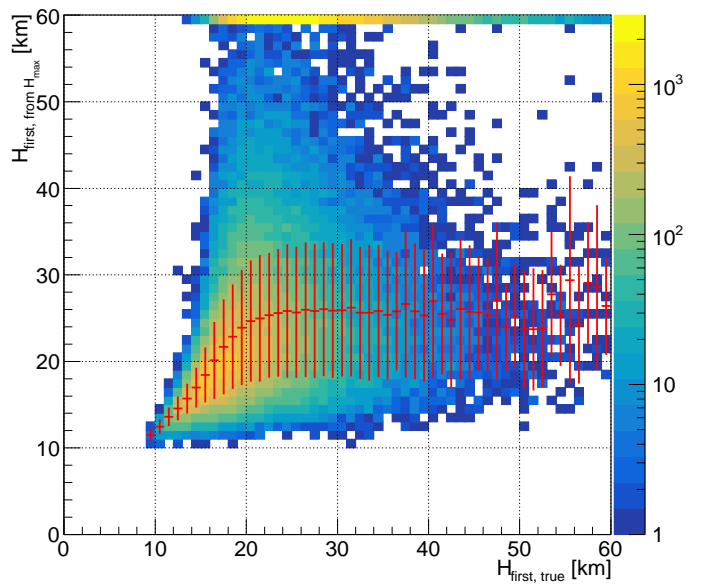}
\caption{Height of the first interaction reconstructed using the height of the shower maximum ($H_\mathrm{max}$) vs the true value of the height of the first interaction for primary gamma rays at the azimuth angle of $180^\circ$.
  Misreconstructed events (with negative reconstructed first interaction depth) are stored in the highest Hmax bin. 
  Gamma-like events with $\theta<0.1^\circ$, that were obtained from the simulated sample of the primary gamma ray at $180^\circ$ azimuth, are plotted in this Figure.
  The red lines show the average and RMS of the reconstructed first interaction depth as a function of the true value of $H_{\rm first}$ (excluding the highest bin with misreconstructed events). 
}
\label{fig:fromxmax}
\end{figure}

\begin{figure}[t]
\includegraphics[width=0.49\textwidth]{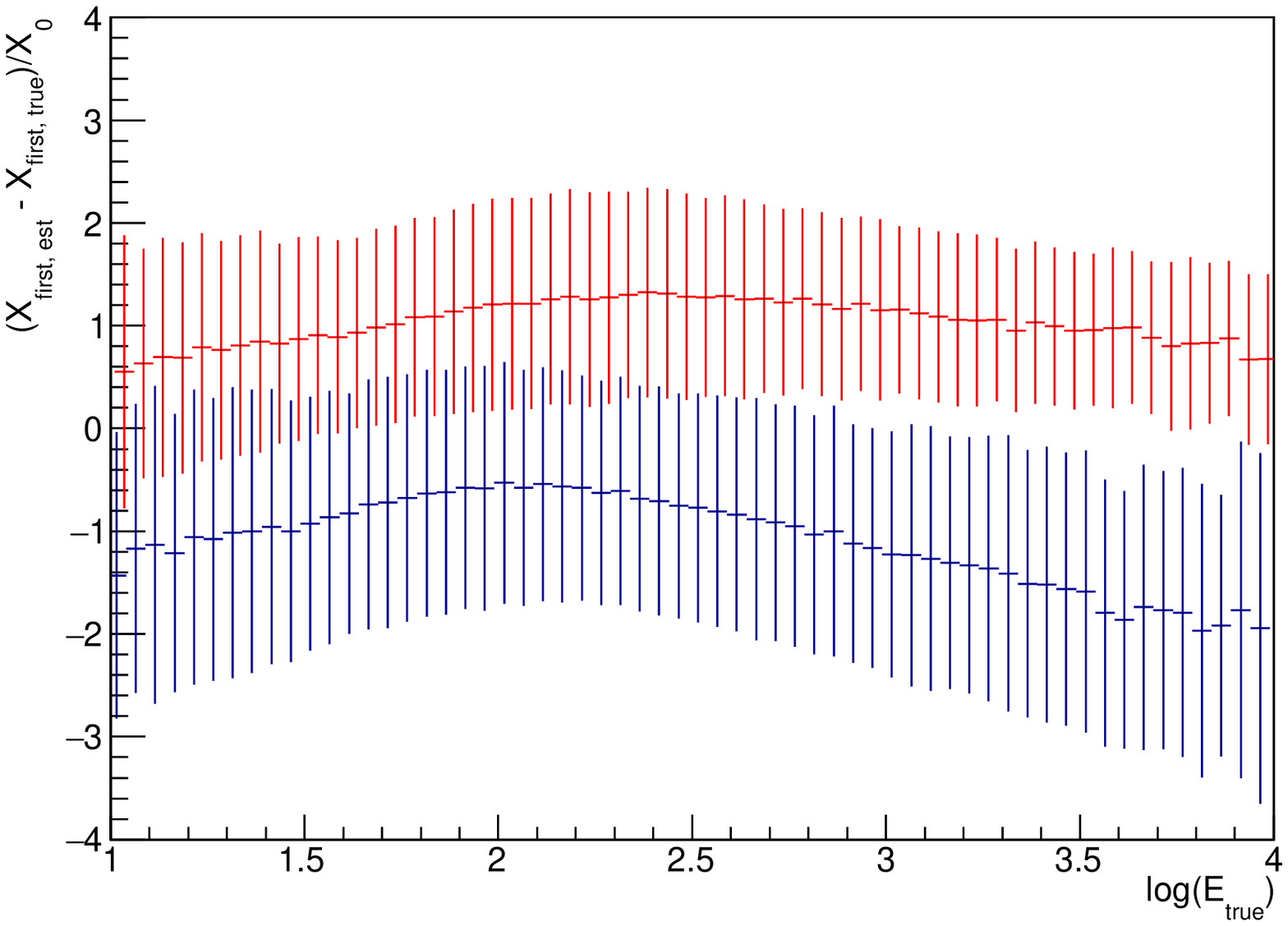}
\includegraphics[width=0.49\textwidth]{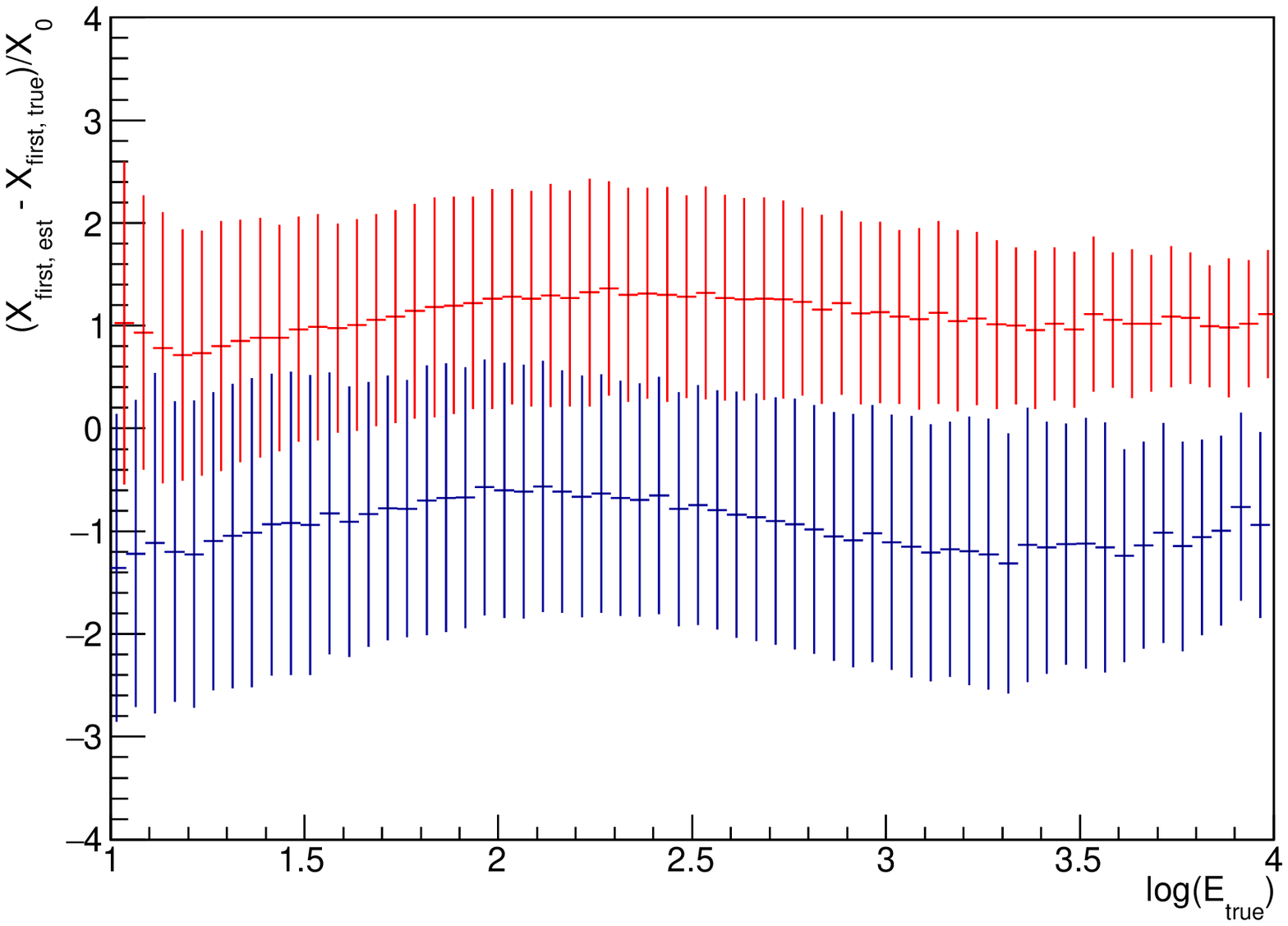}
\caption{Difference between the reconstructed depth of the first interaction and the true value for the $H_\mathrm{max}$ (blue) and the closest pixel (red) methods expressed in radiation lengths.
Only gamma-like events with $\theta<0.1^\circ$ are shown.  
The mean value shows the bias of the method as a function of the energy, and the error bars show the 1-$\sigma$ resolution for primary gamma rays with azimuth angle $0^\circ$ (top panel) $180^\circ$ (bottom panel). 
}
\label{fig:hfirst}
\end{figure}

The correlation of the height of the first interaction reconstructed according to this method with its true value is shown in Fig.~\ref{fig:fromxmax}.
For the correlation plots we selected events that are properly reconstructed i.e. the angular distance, $\theta$, between the reconstructed and the nominal source position is below $0.1^\circ$. 
Moreover, the events  have to be gamma-like, i.e. survive an energy-dependent \emph{Hadronness} cut that preserves 80\% of the gamma ray events (see analysis 1. in Section~\ref{sec:sens}).
The method can only estimate the height of the first interaction for events in which it is  $\lesssim21$\,km (corresponding to about $\mathrm{50\,g\,cm^{-2}}$, similar to the mean interaction path for the gamma ray).
  For showers that start higher in the atmosphere the exclusion of a large fraction of discussed earlier events with estimated negative $X_{\mathrm{1st}}$ produces a strong bias (see red lines in Fig.~\ref{fig:fromxmax}). 
The resolution of this method (including the events from which reconstructed $X_{\mathrm{1st}}$ was negative) is about 0.9--1.5$X_{0}$ with a $\sim -1X_{0}$ bias (see Fig.~\ref{fig:hfirst}).
At the lowest energies ($\lesssim200$\,GeV) the resolution of the method is slightly better ($\sim5$\%, corresponding to $\sim0.06X_0$) for Azimuth angle $0^\circ$ than for $180^\circ$.

\subsection{Using closest pixel}\label{sec:closest}
The second method that we have derived for the estimation of the height of the first interaction is based on the shape of the shower image.
The image of the shower is the two dimensional angular distribution of the Cherenkov light recorded by the telescope.
Thus the image shape is determined by the geometry of the shower, the distance between the shower core position and the telescope and angular distribution of charged particles in the shower folded with height-dependent Cherenkov angle. 
The pixels closest to the direction of the source on the camera carry the information about the top part of the shower. 
As the primary gamma ray does not induce Cherenkov light, its production can start only after the first $e^+e^-$ pair production. 
Hence, we use the angular distance, $d$, between the closest pixel and the reconstructed source position to calculate the height of the first interaction.
To take into account the pixelization of the camera, we roughly correct the distance $d$ adding to it in quadrature the pixel size $d_\mathrm{pix}$ (empirically selected value), i. e.  $d'\!=\!\sqrt{d^2+d_\mathrm{pix}^2}$.  
Next, using the reconstructed impact parameter and the corrected distance $d'$, we compute the height of the first interaction for a single telescope. 
The final value of the height of the first interaction is obtained from averaging individual values from telescopes that fulfill two conditions: $d'\!>\!1.5d_\mathrm{pix}$ (empirically selected value) and the reconstructed impact parameter to the telescope is $>20$\,m.
The images with a small impact value are rejected as they produce images around the true source position. 
This method has additional caveats. Depending on the geometry of the shower, the Cherenkov photons from the topmost part of the shower may do not reach the location at which the telescopes are located.
This effect can be partially counteracted by using only images for which $d'$ is smaller than the Cherenkov angle at the first interaction height estimated from that image.
Such a cut lowers slightly the bias of the method, however prevents estimation of the height of the first interaction for $\sim 30\%$ of the events, hence we do not apply it in the rest of the analysis. 
Also the amount of light from the first $e^+e^-$ pair reaching the telescopes might be too small to produce a signal in a pixel that survives the cleaning.

\begin{figure}[t]
\includegraphics[width=0.49\textwidth]{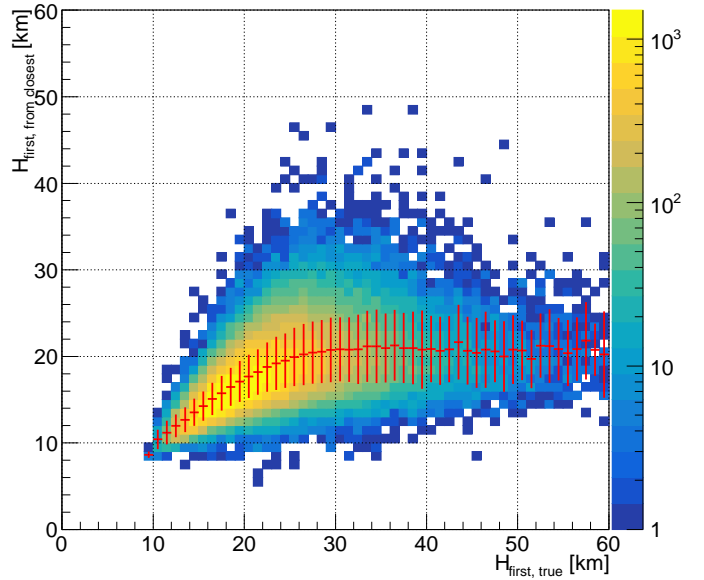}
\caption{Height of the first interaction reconstructed using the closest pixel information vs the true value of the height of the first interaction for primary gamma rays at the azimuth angle of $180^\circ$. 
  Only gamma-like events with $\theta<0.1^\circ$  are shown.
    The red lines show the average and RMS of the reconstructed first interaction depth as a function of the true value of $H_{\rm first}$.
}
\label{fig:fromclosest}
\end{figure}

The correlation of the height of the first interaction reconstructed with this method with its true value is shown in Fig.~\ref{fig:fromclosest}.
As in the case of the method presented in Section~\ref{sec:hmax} the correlation plot contains only gamma-like events with  $\theta<0.1^\circ$.
This method can only estimate the height of the first interaction up to  $\sim30$\,km (corresponding to about $\mathrm{12 \,g\,cm^{-2}}$), above which value a strong bias is seen.
  We note however that only a small fraction of events starts that early in the atmosphere ($\sim 18\%$), and moreover events starting very high in the atmosphere have a smaller chance to trigger the telescopes.
  Hence, we conclude that the method can provide an estimation of the height of the first interaction for most of the events. 
The corresponding resolution is about 0.7--1.3$X_{0}$ with a $\sim 1X_{0}$ bias (see Fig.~\ref{fig:hfirst}). 
The bias is most probably caused by the fact that high energy showers can trigger more distant telescopes.
Hence it is possible that the light from the top parts of the shower does not reach the telescope's position due to the angle at which the light is emitted. 
Such an effect would produce underestimation of the height of the first interaction, as observed in Fig.~\ref{fig:hfirst}. 

In Fig.~\ref{fig:hfirst} we compare the performance of the two methods at different energies of gamma rays for both considered Azimuth angles.
The closest pixel method has a smaller spread of the estimated depth of the first interaction and a bias similar in magnitude (but opposite in direction) to the $H_{max}$ method.
It is interesting to compare the performance for the here presented closest pixel  method also to the model analysis approach \citep{bo98}, where shower images are fitted into templates dependent on the physical parameters of the shower.
The resolution of the depth of the first interaction obtained with the closest pixel method, $(\sim 1 X_{0})$, is of the order of the resolution claimed for the H.E.S.S. telescopes of $\sim0.7 X_{0}$ \citep{nr09}, the latter method has also a significantly smaller bias.
Following \cite{nr09}, we applied a cut of $\theta<0.1^\circ$.
  Additionally, the referenced model analysis method uses a shape goodness quality cut, that conserves 70\% of gamma rays, while for the presented here closest pixel method a \emph{Hadronness} cut conserving 80\% of gamma rays was applied. 
The model analysis requires starting values for all the model parameters to avoid finding local minimum and to improve speed of the algorithm.
Hence, the method presented here can be used to provide such values, possibly improving the performance of the model analysis method.

Finally, the obtained resolution of the depth of the first interaction is also sufficient to provide partial separation of Single Electromagnetic Subcascade (SES, see e.g. \citealp{si18}) events from gamma rays.
Compared to gamma ray showers, the start of SES events should be deeper in the atmosphere by the average value of the proton's first interaction $\sim 90\,\mathrm{g\,cm^{-2}}\approx 2.4 X_0$ (see e.g. \citealp{mi94}).

\section{Results}\label{sec:res}
The new parameters can be used in the analysis of LST data to possibly improve the performance of the telescopes arrays. 
The most natural application is the gamma/background separation procedure which potentially can profit from the knowledge of the height of the first interaction. 
For the gamma/background separation we use the Random Forest (RF) method implemented in \progname{Chimp}/\progname{MARS} \citep{al08}.
For each telescope an individual value of the \emph{Hadronness} is calculated from image and stereo parameters.
The global \emph{Hadronness} value is computed as a weighted average of the values from individual telescopes (see \citealp{si18} for details).
This method allows us to add new parameters and combinations of such parameters into the gamma/background separation method. 

In our analysis we divide the gamma ray MC sample into 4 subsamples: A, B, C and D, and the background sample into 3 subsamples B, C, D.
Subsample gamma-A is used for training the energy estimation. 
We apply the stereo parameters and energy reconstruction to samples gamma-B and background-B, calculate the new parameters for each event and train RF for gamma/background separation.
Next, we apply all the above to samples gamma-C, gamma-D, background-C, background-D.
Subsamples C are used to find the best gamma/background separation cuts for each estimated energy.
Finally, those optimized cuts are applied on subsample D to estimate the final performance in an unbiased way.
The subsamples and their application is summarized in Table~\ref{tab:samples}.
\begin{table}[t]
  \begin{tabular}{|c|c|c|c|}
    \hline
    Subsample & Size & Usage \\\hline\hline
    gamma-A & 1/4 & training energy estimation \\\hline
    gamma-B & 1/4 & \multirow{2}{*}{training g/h separation} \\\cline{1-2}
    background-B & 1/3 &\\\hline    
    gamma-C & 1/4 & \multirow{2}{*}{finding best g/h separation cuts} \\\cline{1-2}
    background-C & 1/3 &\\\hline    
    gamma-D & 1/4 & \multirow{2}{*}{calculate sensitivity} \\\cline{1-2}
    background-D & 1/3 &\\\hline    
  \end{tabular}
  \caption{Summary of the MC subsamples used in the analysis. The individual columns are: tag of the sample, fraction of the full MC sample of a given type used in a subsample, and where it is used in the analysis chain.}
  \label{tab:samples}
\end{table}

\subsection{Low-energy sensitivity improvement}\label{sec:sens}
The CTA sensitivity (and LST in particular)  at the lowest energies ($\lesssim 100$\,GeV) will be mostly limited by the so-called \emph{false gamma} events.
In those most of the primary proton energy is transmitted into a  single electromagnetic cascade (hereafter SES) that induces Cherenkov radiation observed by the telescopes \citep[see][]{maier,so07,so15a,si18}.
Such events have very similar image shapes to real gamma rays, and can be rejected only partially based on the fact that they on average develop slightly deeper in the atmosphere \citep{so15a}. 
In fact \cite{si18} shows that the height of the shower maximum, which is already used in CTA analysis exploits most of the possible separation of such events.
In the case of the LST subarray, the expected improvement of the quality factor is of the order of 15\% only, when the cut in the shower maximum is used to separate gamma rays from SES events. 

In order to study if the estimations of the height of the first interactions proposed in this paper can provide any further improvement in the sensitivity we prepared a set of RFs using different sets of parameters:
\begin{enumerate}
\item \textbf{standard}: image parameters (the Width, Length, Size, Concentration), stereo parameters (the height of the shower maximum, impact parameter) and event-wise and telescope-wise estimation of the energy
\item \textbf{closest pixel}: as in 1., but including the telescope-wise information of the distance between the reconstructed event direction and the closest pixel ($d'$)
\item \textbf{$H_{\mathrm{first}}$ from closest}: as in 1., but including estimation of height of the first interaction from the closest pixel
\item \textbf{$H_{\mathrm{first}}$ from $H_{\mathrm{max}}$}: as in 1., but including estimation of height of the first interaction from the height of the shower maximum
\item \textbf{all new}: all the parameters from 1. -- 4.
\item \textbf{first interaction}: as in 1., but including also the information from \progname{CORSIKA} about the true height of the first interaction.
\end{enumerate}
We are interested mostly in the possible improvement of the sensitivity below 100\,GeV, where the contribution of cosmic ray electrons in the residual background is not so large (see e.g. \citealp{si18}).
Therefore we compare simulated gamma rays against protons only, hence the sensitivities presented here are likely to be overly optimistic in the energy range above a few hundred GeV, where strong contribution of electrons is expected. 
It should be noted that the set 6. of the parameters cannot be used in the real data analysis. 
We test it to estimate the possible gain that the perfectly known height of the first interaction can give us and compare it with the gain obtained with estimated values of the height of the first interaction. 

Note that for protons, the depth of the first interaction obtained from \progname{CORSIKA} is the depth at which the first pions are typically generated, i.e. $\approx 2.4 X_0$ \citep{mi94}. 
But the SES event starts in average $\sim 9/7 X_{0} = 47\,\mathrm{g cm^{-2}}$ later, when e.g. the two gamma rays from $\pi^0$ convert to $e^+e^-$ pairs. 
On the other hand the estimated height of the first interaction discussed in this paper is sensitive to the start of electromagnetic subshower, hence in principle might provide a somewhat better separation than true height of the first interaction.

\begin{figure*}[t!]
\includegraphics[width=0.49\textwidth, trim=0 0 40 15, clip]{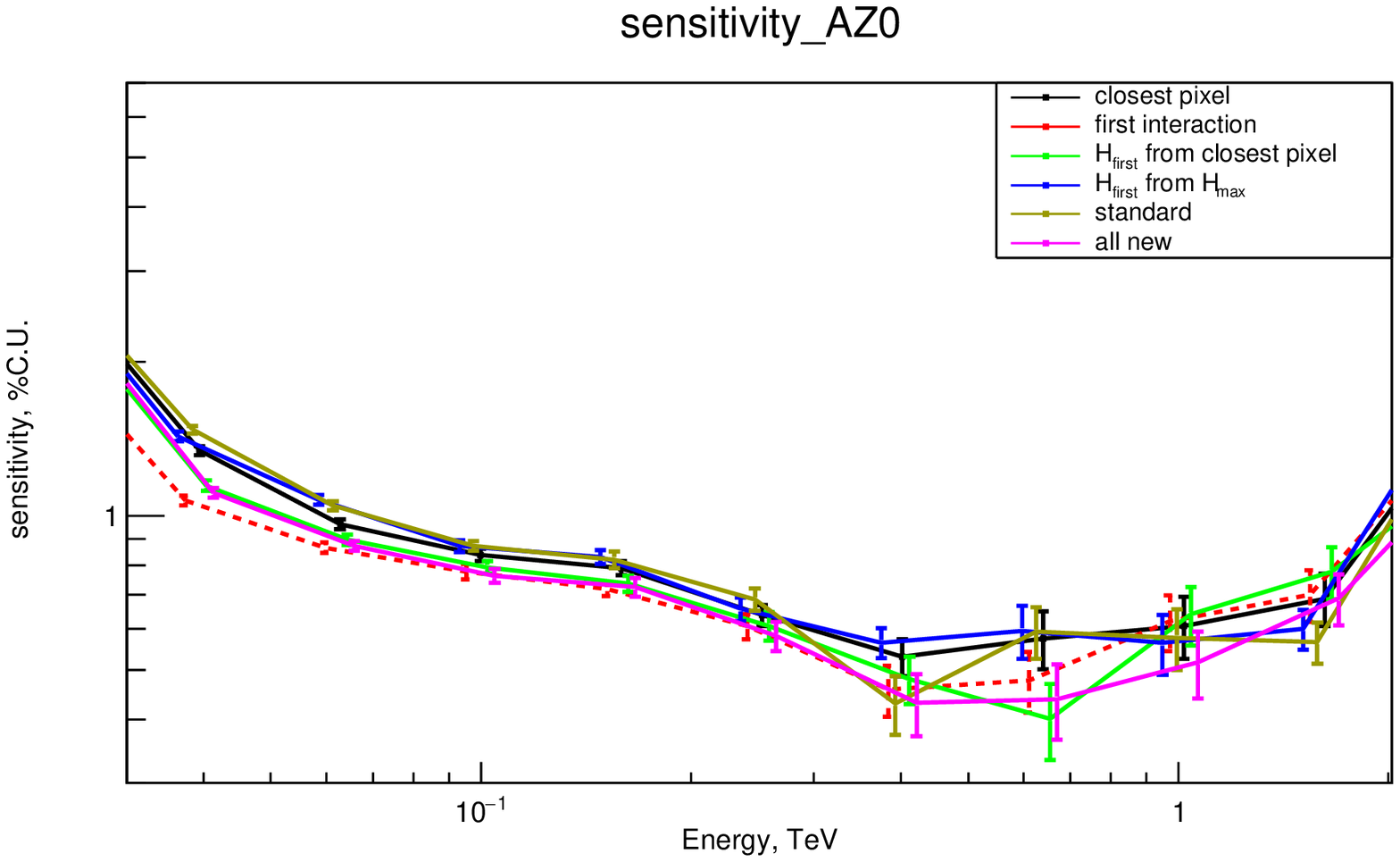}
\includegraphics[width=0.49\textwidth, trim=0 0 40 15, clip]{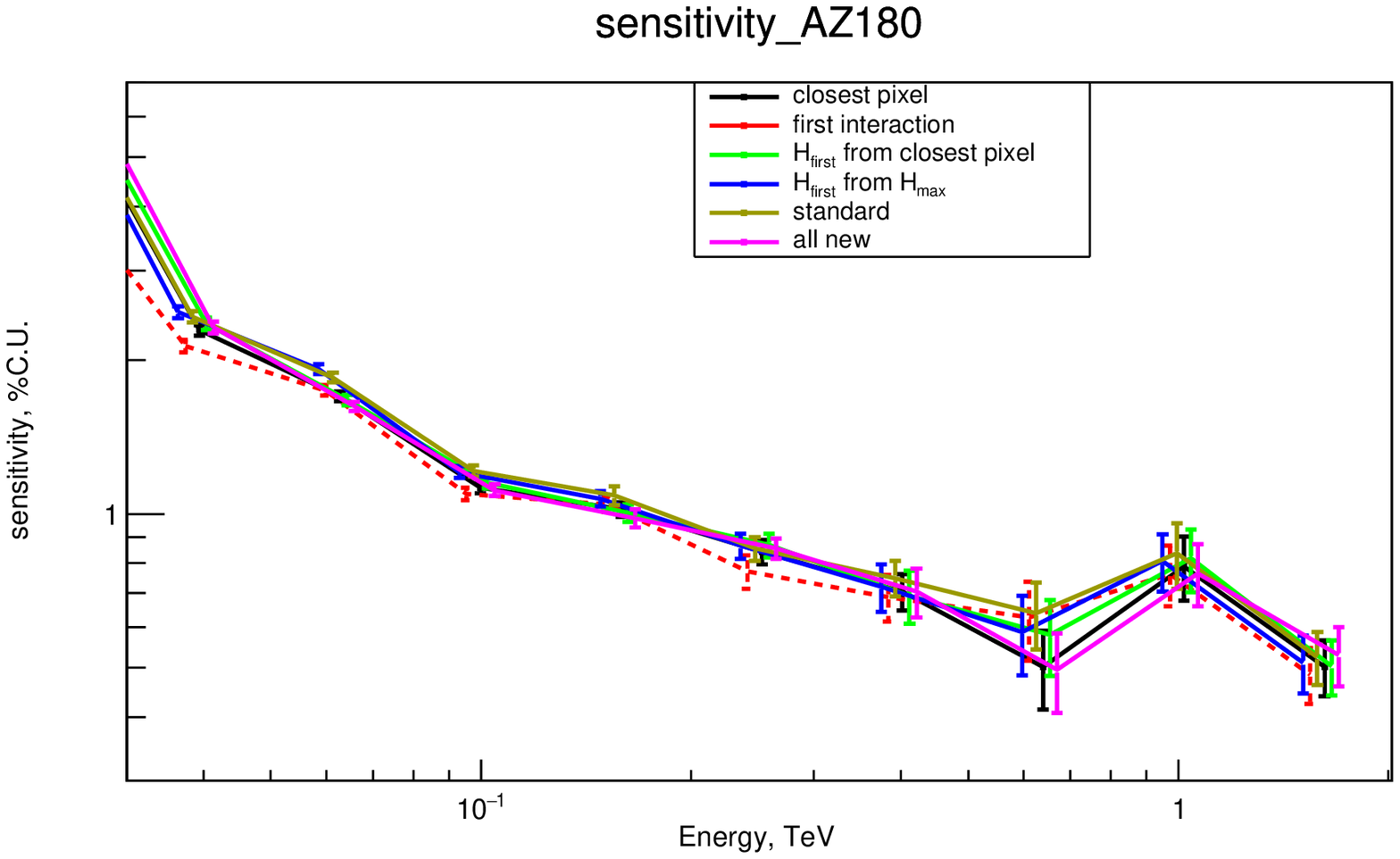}\\
\includegraphics[width=0.49\textwidth, trim=0 0 40 30, clip]{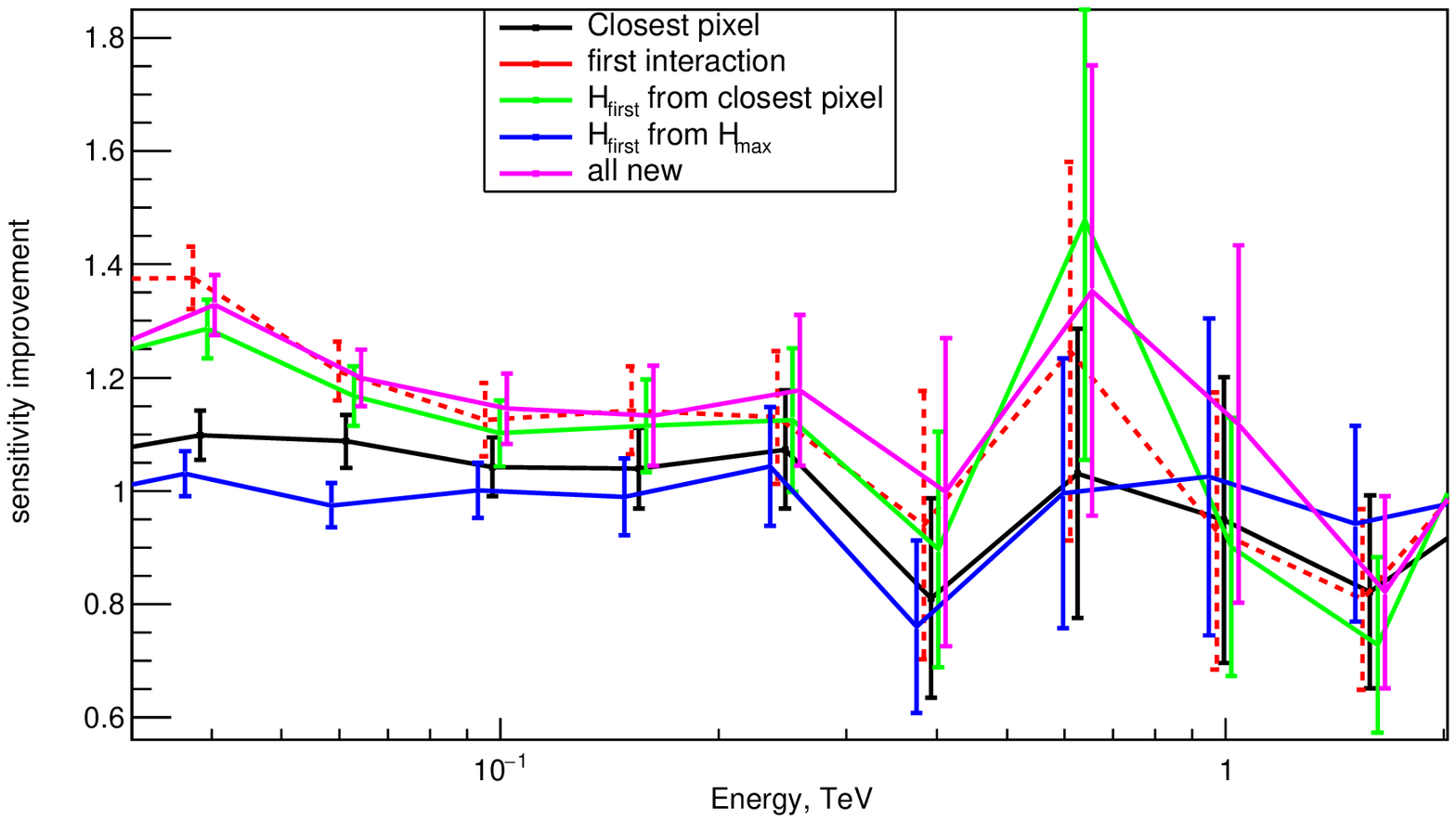}
\includegraphics[width=0.49\textwidth, trim=0 0 40 30, clip]{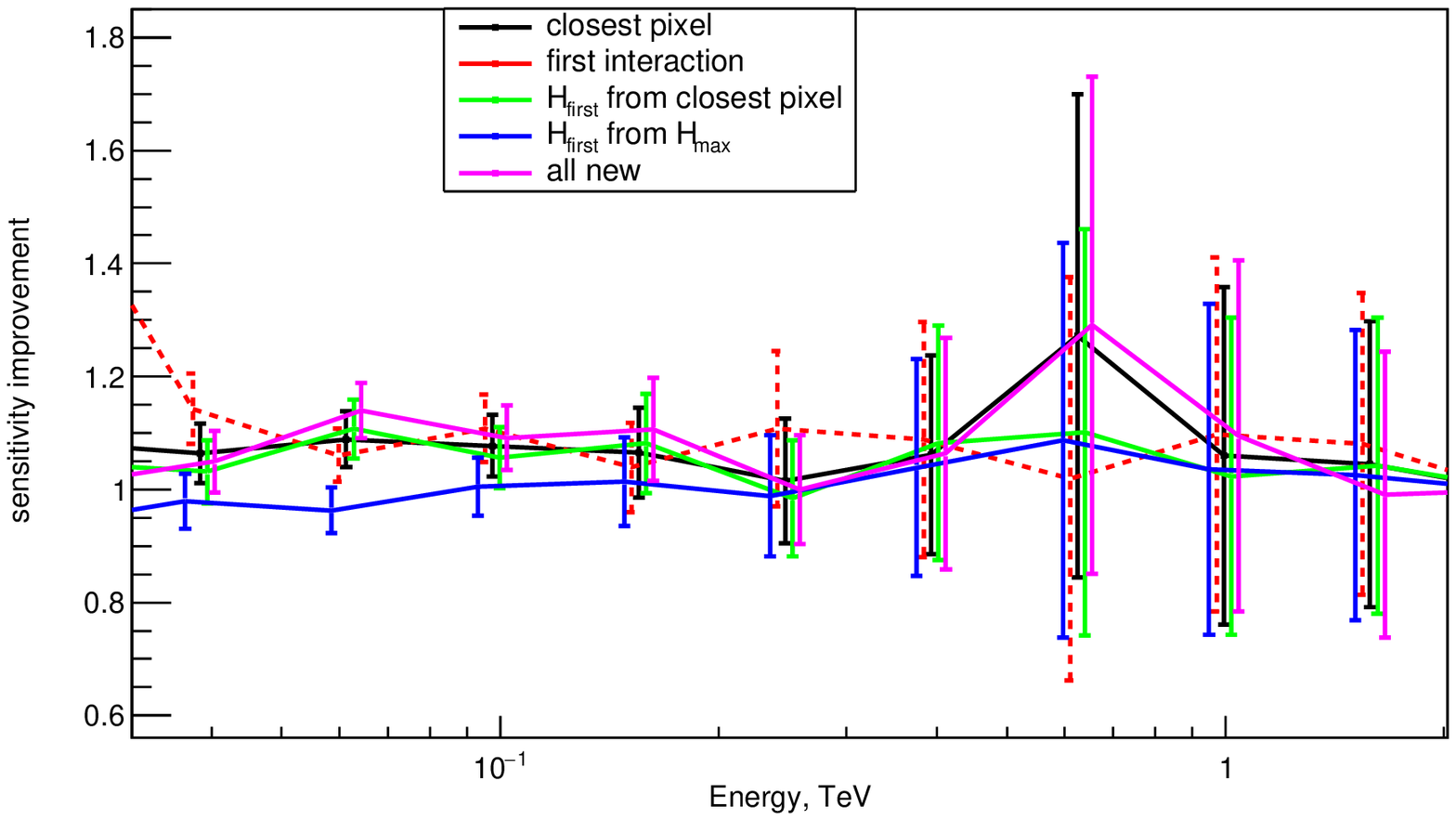}
\caption{Top panels: comparison of differential sensitivities (and their uncertainties) for 50\,hrs observations of 4 LST subarray in CTA North Observatory, obtained with different training parameters (see Section~\ref{sec:sens}): 1. (olive green), 2. (black), 3. (green), 4. (blue), 5. (magenta) and 6. (red dashed). 
Bottom panels: sensitivity improvement with respect to the standard one, 1. (i.e. ratio of sensitivity for 1. analysis and for a given analysis) obtained with different sets of training parameters.
Left and right panels correspond to the azimuth angles $0^\circ$ and $180^\circ$ respectively.
For better visibility of the plot the individual curves are slightly shifted in the X direction.
}
\label{fig:sens}
\end{figure*}

The optimized sensitivities obtained using analyzes 1. -- 6. are presented in Fig.~\ref{fig:sens}. 
The estimation of the height of the first interaction from the $H_{\mathrm{max}}$ (parameter set 4.) does not result in any significant improvement of the sensitivity. 
It is understandable as it is based on the same parameters (i.e. the height of the shower maximum and the reconstructed energy) that are already available in the standard RF. 
Interestingly, a slight improvement in sensitivity below $\sim$100\,GeV was obtained while the parameter of distance of the closest pixel is used (the parameter set 2.). 
The improvement is $\sim10-15\%$ for both tested Azimuth angles. 
If instead the height of the first interaction estimated based on such a distance (parameter set 2) is used than a similar $\sim10\%$ improvement is seen for Azimuth $180^\circ$ (almost perpendicular to the GF), while for Azimuth $0^\circ$ (at a small angle to the GF) the improvement is larger and reaches $\sim 20\%$.
Note, that the sensitivity of an LST calculated here for a 50\,hrs observation time below $\sim 100$\,GeV is rather limited by the condition that the gamma ray excess is above 5\% of the residual background than by the significance of $5\sigma$.
  This means that in this energy range an improvement of 20\% in sensitivity is equivalent to 20\% stronger background rejection.
  In such a case the improvement of the Quality Factor (ratio of the fraction of gamma rays surviving the cut to the square root of the fraction of the background events surviving it) is $\sim 1.1$. 
Usage of all the new parameters (parameter set 5.) give a slightly better performance to the one obtained with adding only the height of the first interaction from the closest pixel. 
The gain in the sensitivity can be compared with the one obtained from the model analysis, which includes also an estimation of the height of the first interaction.
\cite{nr09} showed that an explicit cut in the reconstructed depth of the first interaction brings a small improvement of quality factor
of  1.07. 
  Note that the full gain of the model analysis, which includes the likelihood fitting of the full images, with respect to the standard Hillas-parameters based analysis reported in \cite{nr09} is a factor $\sim 2$.
  This gain might also include partially the dependence of the shower image on the depth of the first interaction reflected in the goodness of the model analysis fit.

It is seen in Fig.~\ref{fig:sens} that the improvement due to the new parameters is more pronounced in the case of Azimuth $0^\circ$, i.e. where the effect of the GF is weaker and the general performance of the array is better.
At the energies where the improvements in sensitivity is seen, the resolution of the first interaction depth is slightly (5-15\%) better for the Azimuth $0^\circ$ case (see Fig.~\ref{fig:hfirst}).
Moreover, the method proposed here for the estimation of the first interaction height exploits the usage of parameters that are affected by the GF: the reconstructed source direction and impact parameter.
Therefore the performance of the method is also affected by the GF. 
In addition, the East-West elongation of images (see e.g. \citealp{co08}), can affect the closest pixel estimation if the shower axis and telescope position are also aligned close to the East-West line.

It is interesting to note that the gain obtained with the usage of the estimated height of the first interaction is similar to the gain from using such a parameter extracted from the \progname{CORSIKA} simulation. 
The sensitivity improvement due to the estimated height of the first interaction (10-20\%) is also of the same order as the claimed possible improvement of the better estimation of the height of the shower maximum ($\sim$15\%, average over Azimuth $0^\circ$ and $180^\circ$, \citealp{si18}). 
As the height of the shower maximum fluctuates more than the height of the first interaction, the separation power of the latter one should be more powerful. 
A toy MC study similar to the one used in \citet{si18} shows that perfect knowledge of the height of the first  pair production of primary gamma rays and SES events would allow us to get 25\% better quality factor than knowledge of the true depth of the shower maximum.

\subsection{Electron rejection}\label{sec:ele}
\begin{figure}[t]
\includegraphics[width=0.49\textwidth]{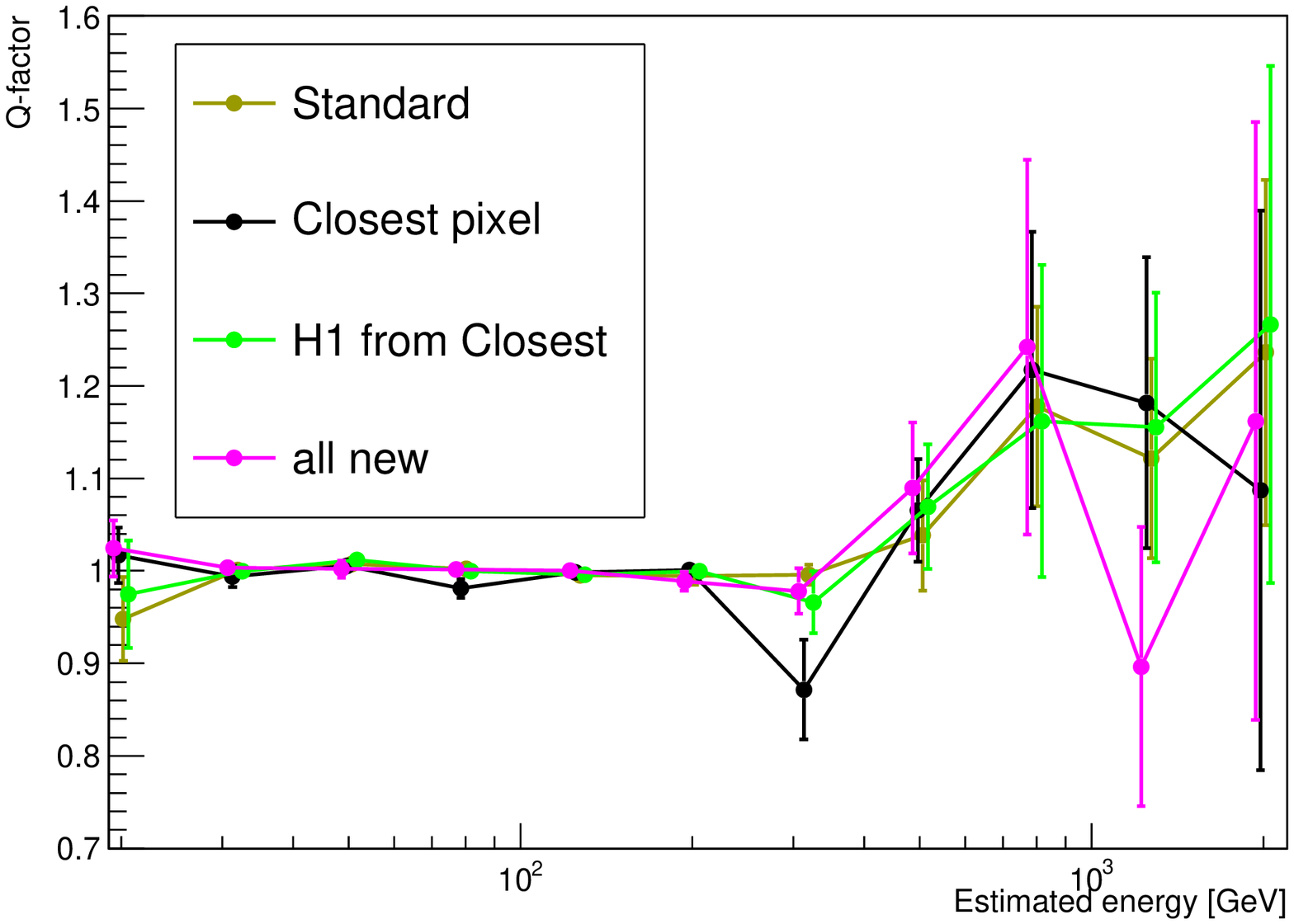}
\includegraphics[width=0.49\textwidth]{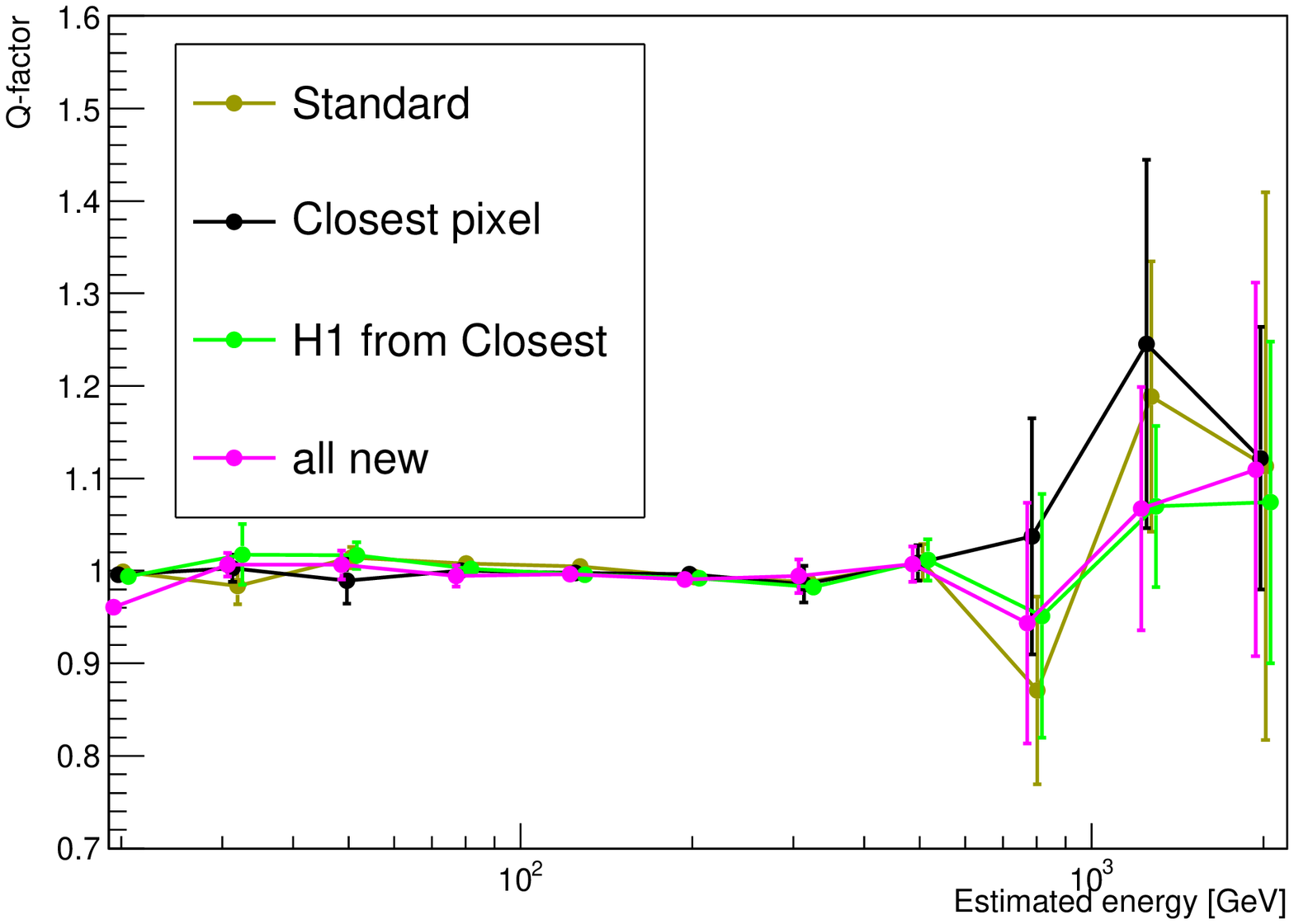}
\caption{Q-factor of the \emph{Electronness} cut between gamma and electron simulations as a function of the estimated energy for Azimuth $0^\circ$ (top) and $180^\circ$ (bottom).
  \emph{Electronness} parameter is obtained with different sets of training parameters (see Section~\ref{sec:ele}): 1. (olive green), 2. (black), 3. (green) and 5. (magenta).
  For better visibility of the plot the individual curves are slightly shifted in the X direction.
}
\label{fig:qfact}
\end{figure}

SES events are not the only type of background that is difficult to separate from gamma rays.
At energies above a few hundred GeV the separation of the hadronic background is very efficient.
However, despite having a softer spectrum, electrons constitute a significant fraction of remaining background \citep[see e.g.][]{pa16,am17,si18}, significantly limiting the sensitivity in the energy range from a few hundred GeV up to a few TeV \citep{acha13}. 
As the electrons produce electromagnetic cascades they are nearly indistinguishable from gamma rays.
Any possible separation might come only from a difference in the first interaction and in the development of the shower in the first generations.
The difference in the mean free path for the pair production of gamma rays ($9/7\ X_0$) and the radiation length $X_0$ at which the electrons radiate most of their energy via Bremsstrahlung is rather subtle.
Assuming that a gamma ray can be considered as equivalent to two $e^{-/+}$ of half its energy we can expect the shift of the shower maximum by $(9/7-\log(2))X_0 \approx 0.6X_0$, much smaller than the $\sim2.4X_0$ effect between gamma rays and SES events (see Section~\ref{sec:sens}). 
Nevertheless, we test if the parameters proposed by us can improve the sensitivity by more efficient gamma/electron separation.
We apply a similar approach like in Section~\ref{sec:sens}, but for the case of gamma/electron separation we test only those RFs: 
\begin{enumerate}
\item \textbf{standard}: as 1. in Section~\ref{sec:sens}
\item \textbf{closest pixel}: as in 1., but including the telescope-wise information of the distance between the reconstructed event direction and the closest pixel
\item \textbf{$H_{\mathrm{first}}$ from closest}: as in 1., but including estimation of height of the first interaction from the closest pixel
  \setcounter{enumi}{4}
\item \textbf{all new}: all the parameters from 1. -- 3.
\end{enumerate}
The electron/gamma separation parameter obtained from these RFs we call \emph{Electronness} in contrast to \emph{Hadronness} parameter used for separation of gamma/proton in Section~\ref{sec:sens}. 
Instead of maximizing the sensitivity we select the \emph{Electronness} cut in each estimated energy bin to provide the best quality factor $Q=f_g/\sqrt{f_e}$, where $f_g$ and $f_e$ are the fractions of gamma rays and electrons respectively surviving the \emph{Electronness} cut. 
To avoid edge effects we select only electron events with reconstructed direction within $0.5^\circ$ from the camera center.
The obtained $Q$-factors are shown in Fig.~\ref{fig:qfact}. 
Even while a cut in \emph{Electronness} can be tuned to exclude a larger fraction of electrons than gamma rays, neither the standard nor the new parameters bring any significant improvement in sensitivity by rejection of electron-induced showers.
This is also in line with the results of the model analysis where a large overlap of the depth of the reconstructed first interaction between electrons and gamma rays is reported \citep{nr09}.

\section{Conclusions}\label{sec:con}
We have investigated two independent methods for estimation of the height of the first interactions in arrays of Cherenkov telescopes.
Despite their simplicity, both methods have resolution of the order of a single radiation length (comparable to the resolution obtained in the model analysis \citealp{nr09}).
As an example, we have applied the two new methods to the simulated data of a CTA North subarray of 4 LST telescopes and studies possible improvement of sensitivity by more efficient background rejection.
 The bias limits the application of the methods based on the height of the shower maximum and on the closest pixel distance to showers starting not higher in the atmosphere than $\sim21$\,km and $\sim 30$\,km, respectively. 
We found that the method based on the extrapolation of the height of the shower maximum does not bring a significant gain if the height of the shower maximum is already used in the gamma/hadron selection RF.
On the other hand the method exploiting the knowledge of the distance between the closest pixel and the reconstructed source position brings a moderate gain of 10--20\% in the sensitivity at the lowest energies ($\lesssim 100$\,GeV).
The above improvement of sensitivity comes solely from the more efficient rejection of hadronic events, as there is no significant gain in sensitivity by more efficient rejection of electron events with the new parameters. 
The improvement of the gamma/hadron separation efficiency obtained by using the estimated height of the first interaction is of the same order as the gain that would come from the knowledge of true height of the first interaction.
Moreover, both numbers are also of the same order as the estimated in \citet{si18} possible improvement in sensitivity obtained with hypothetical perfect reconstruction of the shower maximum height in single electromagnetic subcascade events.
It should be noted however that in hadronic showers the true height of the first interaction registered by \progname{CORSIKA} describes where the pions are generated.
$\pi^0$ will decay nearly instantaneously into gamma rays, however those still need to traverse about $50 \mathrm{g\,cm^{-2}}$ before creating $e^+e^-$ pair that can induce Cherenkov light.
Instead, for gamma rays the first interaction of \progname{CORSIKA} is directly the height at which the $e^+e^-$ pair is created. 
Hence in principle larger separation could be achievable.

Interestingly, the relative improvement of the height of the first interaction estimation in the gamma/hadron separation depends on the Azimuth angle of the observations.
  The improvement is more significant for the case of the weaker geomagnetic field influence, when also the general performance of the instrument is better.
  This can be understood as the accuracy of the method determining the first interaction height from the distance to the closest pixel of the image will naturally depend also on the angular resolution.
  This shows once again that the geomagnetic field causes physical limits on the analysis of low energy data of IACTs.


\section*{Acknowledgements}
This work is supported by the grant through the Polish Narodowe Centrum Nauki No. 2015/19/D/ST9/00616.
DS is supported by the National Science Centre grant No. UMO-2016/22/M/ST9/00583. 
We would like to thank CTA Consortium and MAGIC Collaboration for allowing us to use their software.
The first stage of the simulations was carried out at the Max Planck Institute for Nuclear Physics in Heidelberg.
This paper has gone through internal review by the CTA Consortium. 
We would like to thank in particular G.~Maier, D.~Parsons, D.~Morcuende, E.~de O\~na Wilhelmi, A.~Mitchell and anonymous journal referee for their valuable comments to the manuscript.

\end{document}